\documentclass[showpacs,superscriptaddress,twocolumn,prl]{revtex4} 
\usepackage{amsmath,graphicx}
\usepackage{subfigure}
\usepackage[usenames]{color}
\usepackage[dvipsnames]{xcolor}
\usepackage{ulem}

\begin{document} 
 
\title{Nonadiabatic Dynamics of Strongly Driven Diffusive Josephson Junctions}
\author{J. Basset } 
\affiliation{Laboratoire de Physique des Solides, CNRS, Univ. Paris-Sud,
	Universit\'e Paris Saclay, 91405 Orsay Cedex, France }  
\author{M. Kuzmanovi\'c} 
\affiliation{Laboratoire de Physique des Solides, CNRS, Univ. Paris-Sud,
	Universit\'e Paris Saclay, 91405 Orsay Cedex, France  } 
\author{P. Virtanen}
\affiliation{NEST, Istituto Nanoscienze-CNR and Scuola Normale Superiore, I-56127 Pisa,Italy}
\affiliation{University of Jyv\"askyl\"a, Department of Physics and Nanoscience Center, P.O. Box 35 (YFL), FI-40014 University of Jyv\"askyl\"a, Finland}
\author{T. T. Heikkil\"a}
\affiliation{University of Jyv\"askyl\"a, Department of Physics and Nanoscience Center, P.O. Box 35 (YFL), FI-40014 University of Jyv\"askyl\"a, Finland}
\author{J. Est\`eve} 
\affiliation{Laboratoire de Physique des Solides, CNRS, Univ. Paris-Sud,
	Universit\'e Paris Saclay, 91405 Orsay Cedex, France  } 
\author{J. Gabelli} 
\affiliation{Laboratoire de Physique des Solides, CNRS, Univ. Paris-Sud,
	Universit\'e Paris Saclay, 91405 Orsay Cedex, France  } 
\author{C. Strunk} 
\affiliation{Laboratoire de Physique des Solides, CNRS, Univ. Paris-Sud,
	Universit\'e Paris Saclay, 91405 Orsay Cedex, France } 
\affiliation{Institute of Experimental and Applied Physics, University of Regensburg, D-93040 Regensburg, Germany}
\author{M. Aprili} 
\affiliation{Laboratoire de Physique des Solides, CNRS, Univ. Paris-Sud,
	Universit\'e Paris Saclay, 91405 Orsay Cedex, France  } 
 

\begin{abstract}

By measuring the Josephson emission of a diffusive Superconductor-Normal metal-Superconductor (SNS) junction we access the harmonic content of the current-phase relation (CPR). We experimentally identify a novel non-adiabatic regime in which the CPR is modified by high frequency microwave irradiation. This observation is explained by the excitation of quasiparticles in the normal wire induced by the electromagnetic field.
The distortion of the CPR originates from the phase-dependent out-of-equilibrium distribution function which is strongly affected by the ac-response of the spectral supercurrent. For a phase difference approaching $\pi$, transitions accross the minigap are dynamically favored leading to a supercurrent reduction. This finding is supported by a comparison with the quasiclassical Green's function theory of superconductivity in diffusive SNS junctions under microwave irradiation.

\end{abstract}

\maketitle

 At sufficiently low temperatures, superconductors cannot absorb microwave radiation of energy smaller than the superconducting energy gap $\Delta$ \cite{Glover1956,Mattis1958,TinkhamBook}.  	In Josephson weak links instead, where two superconductors (S) are weakly coupled through a long diffusive metallic wire (N),  radiation can be absorbed in N because the induced gap in the density of states or minigap \cite{Belzig1996,Ferrier2013} is considerably smaller than $\Delta$. In this  Rapid Communication we show that the out-of-equilibrium state originating from such absorption and its feedback on the quasiparticle spectrum of the wire strongly modifies the current-phase relation (CPR) \cite{Golubov2004} of the junction. In particular we observe a large increase of its second harmonic which reflects the peculiar out-of-equilibrium distribution function obtained under high frequency microwave irradiation.  This finding is in good agreement with the quasi-classical theory of superconductivity in which the effect of the microwave drive on the spectral current density is taken into account \cite{Virtanen2010}.
 
In  proximity-coupled Josephson junctions, Andreev reflections lead to a coherent superposition of electron-hole excitations in the weak link, which carry the supercurrent \cite{Yip1998,Heikkila2002}. These excitations form a quasi-continuum of Andreev bound states (ABS)\cite{Ferrier2013,Heikkila2002}.
The single particle density of states  in N develops a minigap $E_{\mathrm{g}}(\varphi)$ whose amplitude depends on the phase difference, $\varphi$, between the two superconductors \cite{Kulik1970,Golubov2004,Lesueur2008} and is minimal for $\varphi=\pi$ \cite{Ivanov2002, Dassonneville2018}. In long wires the minigap is set by the diffusion time $\tau_D = L^2/D$ and is proportional to the Thouless energy, $E_{\mathrm{Th}}= \hbar /\tau_D$ as $E_{\mathrm{g}}(0)\approx3.1E_{\mathrm{Th}}\ll\Delta$ \cite{Zhou1998}, where $D$ and $L$ stand for the diffusion coefficient and the length of the wire, respectively. The supercurrent is related to the Andreev spectrum via the spectral current density $j_{s}(E, \varphi)$ and the distribution function $f(E, \varphi)$ \cite{Yip1998}:
\vspace{-5pt}
\begin{equation} 
\label{CPReq}
I(\varphi)= \frac{1}{e R_{\mathrm{N}}}\int{[1-2f(E,\varphi)] j_{\mathrm{s}}(E,\varphi)dE},
\end{equation}
where $R_{\mathrm{N}}$ is the normal state resistance of the wire.
The periodic phase dependence in $j_{s}(E, \varphi)$ gives rise to a Fourier expansion of $I(\varphi)$ with coefficients $I_{\mathrm{c,n}}$, such that the CPR reads \cite{Heikkila2002}:
\vspace{-5pt}
\begin{equation} 
\label{CPReq2}
I(\varphi)= \sum_{n=1}^{\infty} I_{\mathrm{c,n}} \sin (n \varphi).
\end{equation}
At thermal equilibrium $f(E)$ is the Fermi distribution function, and is independent of $\varphi$.

The purpose of this work is to induce and probe the out-of-equilibrium state obtained in the \textit{strongly non-adiabatic} regime for which the frequency of the microwave drive $\omega_{\mathrm{rf}}$ exceeds both the energy relaxation rate $\Gamma$ and the minigap: $\Gamma<2E_{\mathrm{g}}/\hbar\lesssim\omega_{\mathrm{rf}}$~\cite{note1}. In this novel situation  both  the spectral supercurrent $j_{s}(E, \varphi)$ and the distribution function are altered by the pair-breaking induced by the microwave absorption \textit{i.e.} by a direct excitation of quasiparticles across the minigap.

Experimentally we address $I_{\mathrm{c,n}}$ by measuring the ac-Josephson effect \cite{Langenberg1965} under microwave illumination. We demonstrate that the harmonic content of the Josephson emission is drastically modified due to the quasiparticle energy redistribution within the normal wire. The comparison with the microscopic theory \cite{Virtanen2010} reveals that the time-dependence of the ABS spectrum is essential, as the effect arises from the back-action of the time-dependent spectrum to the out-of-equilibrium distribution function. This observation, in the strongly non-adiabatic regime, goes beyond the usual Eliashberg approximation \cite{Eliashberg1986} in which the ac-spectral supercurrent plays no role \cite{Warlaumont1979,Chiodi2009}.

\begin{figure}[b]
	\begin{center}
		\includegraphics[width=\linewidth]{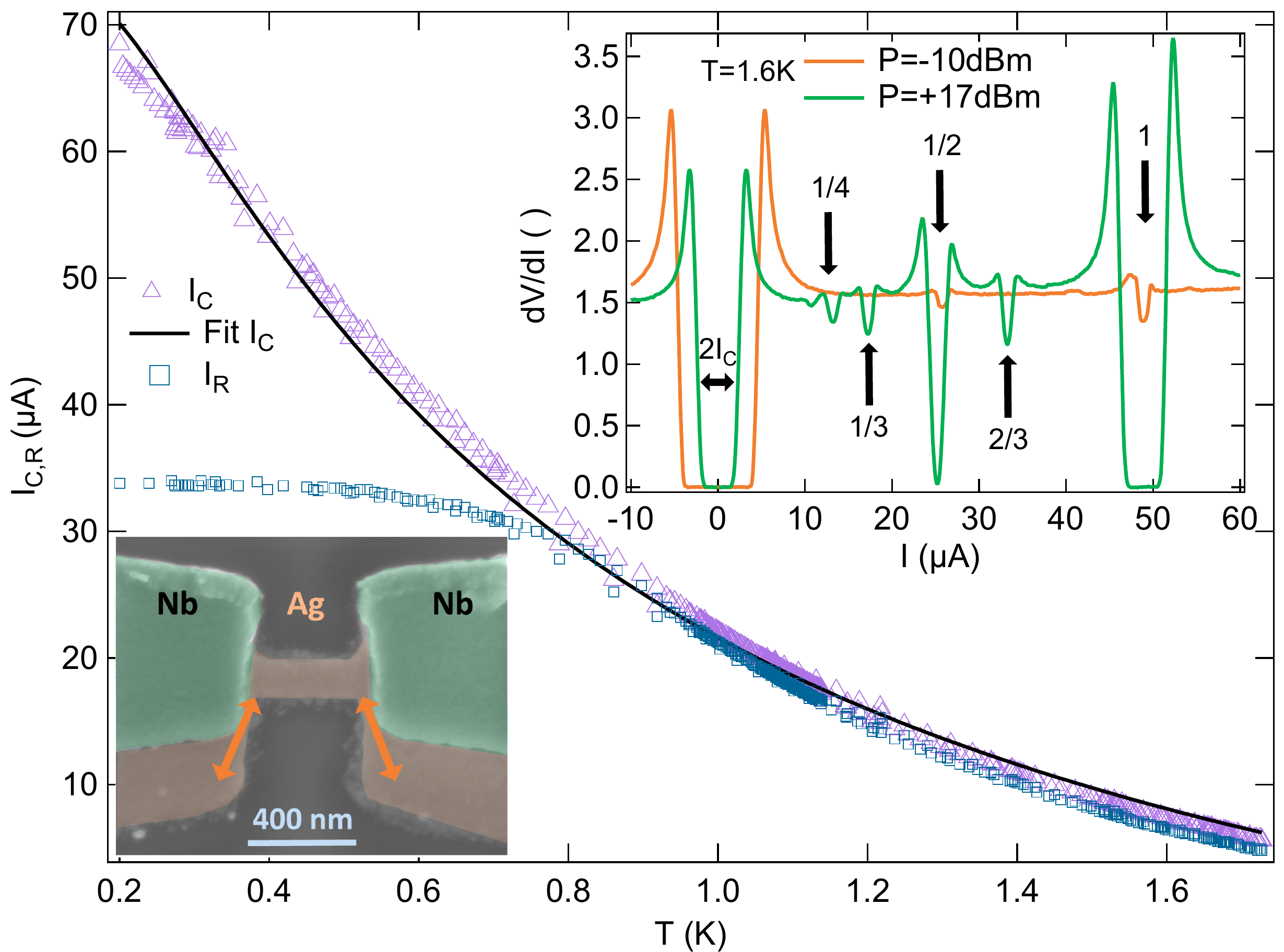}
	\end{center}
	\vspace{-15pt}
	\caption{Temperature dependence of the critical- and retrapping currents of the junction. Top inset: differential resistance $dV/dI$ \textit{vs} dc-current $I$ at $T=1.6$~K for two irradiation powers ($\omega_{\mathrm{rf}}$/2$\pi=35.18$~GHz). The high power curve exhibit subharmonic Shapiro steps (see arrows and corresponding fractions). Bottom inset: scanning electron micrograph of the junction. Green shading highlights the superconductor (Nb), light orange the normal part (Ag). Arrows point at metallic reservoirs acting as heat sinks.}
	\label{Figure1}
\end{figure}

To investigate the ac-Josephson emission, we have fabricated a radio-frequency compatible SNS junction by e-beam lithography. The junction is obtained by angular e-gun evaporation of a $70$~nm thick layer of Nb (S) and a $40$~nm thick layer of silver (N) (see fig.~\ref{Figure1}(b)). The normal metal length is $L=400$~nm and it has a normal state resistance $R_{\mathrm{N}}=1.6~\Omega$. Normal metal reservoirs (see inset of fig.~\ref{Figure1}) act as heat sinks reducing the energy relaxation times of quasiparticles.
The measurement circuit is presented in the supplementary information (S.I.). The sample is connected  through two bias-tees  which allow dc-biasing, microwave excitation ($\omega_{\textrm{rf}}/2\pi \in[0-40]$~GHz) and detection.

The temperature dependence of the critical current $I_{\textrm{c}}(T)$ together with the retrapping current $I_{\textrm{r}}(T)$, are presented in the main panel of figure \ref{Figure1}. The two curves separate below $T_{\textrm{h}}\approx0.8$~K  where self-heating becomes relevant \cite{DeCecco2016}. We fit the $I_{\textrm{c}}$ data (black line in figure \ref{Figure1}) to obtain an estimate of the Thouless energy $E_{\mathrm{Th}}\approx 19\pm 2~\mu$eV \cite{Dubos2001} which sets the minigap to $2E_{\rm{g}}(0)\approx 118~\mu$eV $\equiv$~$28.5$~GHz. 
By comparing with two shorter samples we verified that the Thouless energy scales as $1/L^2$ provided that the effective wire length is roughly $250$~nm longer than the geometrical gaps between the Nb leads as observed in previous experiments (see S.I. and \cite{Fuechsle2009, Dubos2001, Dubos2001b}). Finally the diffusion coefficient is  found to be $D\approx90$~$\textrm{cm}^2/\textrm{s}$ (see S.I. which also include Refs.~\cite{Likharev1979,Pierre2003,Steinbach1996,Isaacs1965,Ingold1992}) which is close to previous experiments using similar junctions \cite{Fuechsle2009}.
The inset of figure \ref{Figure1} shows the differential resistance as a function of the dc-current bias under microwave excitation ($\omega_{\mathrm{rf}}/2\pi=35.18$~GHz) at $T\approx1.6$~K~$>T_{\textrm{h}}$. The zero resistance plateaus correspond to Shapiro steps at $V_{\mathrm{dc}}=n/m~\hbar\omega_{\mathrm{rf}}/2e$ ($n$ and $m$ integers) \cite{Shapiro1963}. The temperature dependence of the maximum amplitude $I_{\textrm{S}}$ of the main Shapiro step ($n=1,m=1$) allows to verify the quality of the heat sinks (see Ref.~\onlinecite{Basset2019bis} and S.I.)  and deduce the quasiparticle energy relaxation rate $\Gamma/2\pi\approx 4.6$~GHz which corresponds to the escape time of the hot quasiparticles out-of the junction given by the diffusion time $\rm \tau_D=1/\Gamma\approx 35$~ps.
\begin{figure*}[htbp]
	\begin{center}
		\includegraphics[width=1.0\textwidth]{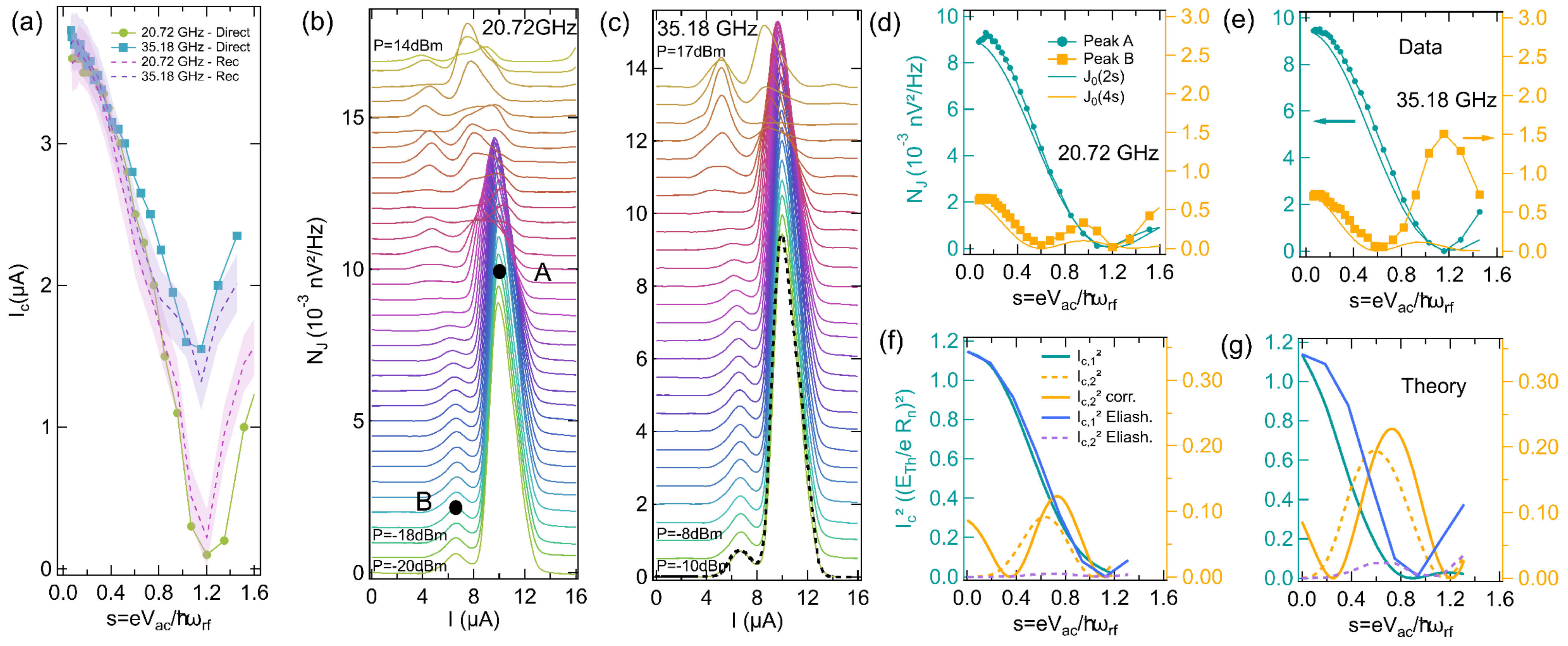}
	\end{center}
	\vspace{-15pt}
	\caption{(a) Power dependence of the critical current for two applied microwave frequencies together with reconstructed critical currents (see text). (b) Ac-Josephson spectral density $N_{\textrm{J}}$ \textit{vs} dc-current for increasing microwave power (The powers expressed in dBm are the one at the output of the microwave generator. The power step size between curves is 1dBm) at $\omega_{\mathrm{rf}}/2\pi=20.72$~GHz (c) Idem (b) for $35.18$~GHz. Dashed curve is the expected emission within the $\omega_0$ band (see S.I.). (d) Power dependence of emission peak amplitudes A and B in figure \ref{Figure2}(b) for $\omega_{\mathrm{rf}}/2\pi=20.72$~GHz. Bessel functions $J_0$ correspond to the adiabatic limit (see text) and are scaled to match the lowest power data points (e) Idem (d) for $35.18$~GHz. (f) Calculated power dependence of the squared harmonics $I_{\textrm{c,1}}^2$ and $I_{\textrm{c,2}}^2$ (proportional to experimental $N_{\textrm{J}}$) for parameters $\hbar\omega_{\textrm{rf}}/E_{\mathrm{Th}}=3$, $\Gamma/E_{\mathrm{Th}}=0.4$, $k_{\mathrm{B}} T/E_{\mathrm{Th}}=7$ and $\Delta/E_{\mathrm{Th}}=55$. (g) Idem (f) but for $\hbar\omega_{\textrm{rf}}/E_{\mathrm{Th}}=7$.}
	\label{Figure2}
\end{figure*}
To further characterize our junction we show in figure \ref{Figure2}(a) the critical current \cite{note0} as a function of the normalized applied microwave field amplitude $s=e V_{\mathrm{ac}}/\hbar \omega_{\mathrm{rf}}$ for two excitation frequencies $\omega_{\mathrm{rf}}/2\pi=20.72$~GHz and $\omega_{\mathrm{rf}}/2\pi=35.18$~GHz. As one increases the microwave power the critical current follows roughly the zero-th order Bessel function $|J_0(2s)|$. Note that the absolute value of $s$ is hard to calibrate accurately. We have here chosen to scale $s$ such that the minimum of the experimental data ($I_{\textrm{c}}$) and the minimum of $|J_{0}(2s)|$ (adiabatic limit) match. Interestingly, the critical current $I_{\textrm{c}}$ for $\omega_{\mathrm{rf}}/2\pi=35.18$~GHz does not vanish at $s\approx 1.2$ as expected in the adiabatic limit \cite{LikharevBook,BaroneBook}. We address this new regime by analyzing the CPR.

The CPR of long SNS junctions under microwave radiation has been investigated in Ref.~\onlinecite{Fuechsle2009} in a phase-biased configuration using a Hall sensor and low microwave frequencies ($\omega_{\mathrm{rf}}<2E_{\rm{g}}/\hbar$). The alternative approach we take in this experiment is to directly measure the ac-Josephson emission spectral density $N_{\textrm{J}}$~($\textrm{V}^2/\textrm{Hz}$) generated by the junction when dc-current biased across a microwave circuit allowing a galvanic coupling to microwaves. We perform the experiment in the limit where the Josephson frequency is small compared to the excitation frequency ($\omega_{\mathrm{J}}=2 e V_{\mathrm{dc}}/\hbar<\omega_{\mathrm{rf}}$) so that the two frequency scales are separated and we can consider a modified CPR with the fast oscillation averaged out (see S.I. for details).  
The frequency of the emitted ac-radiation from the $n^{th}$ harmonic of the CPR obeys the relation $\omega_{\mathrm{J,n}}/2\pi=2enV_{\mathrm{dc}}/h$. Therefore at a fixed dc voltage the harmonic content of the CPR appears as multiple peaks in the spectrum of the emitted Josephson radiation. As it is technically very demanding to perform such an experiment in a large bandwidth,  we adopted a strategy in which the radiation is measured in a band of about $2.5$~GHz centered around $\omega_0/2\pi=6.5$~GHz. In this experimental situation, the contribution from the $n^{th}$ harmonic appears as a radiation peak when the voltage is equal to $V_{\textrm{dc,n}}=\hbar\omega_0/2en$.

We then measure the Josephson radiation spectral density $N_{\textrm{J}}$ as a function of the applied dc-current and microwave power for different $\omega_{\mathrm{rf}}$~\cite{note2}. Such measurements, presented in figure \ref{Figure2}(b) and (c), show two emission peaks at $V_{\mathrm{dc}}\approx\hbar \omega_0/4e\approx6~\mu$eV ($I_{\mathrm{dc}}=6.5~\mu$A at low power) and $V_{\mathrm{dc}}\approx\hbar \omega_0/2e\approx12~\mu$eV ($I_{\mathrm{dc}}=10~\mu$A at low power) corresponding respectively to the second and the first harmonic of the CPR  (letter \textbf{B} and \textbf{A} in figure \ref{Figure2}(b)).
The width of these two peaks is set by the combined effects of thermal noise and the finite measurement bandwidth of the setup (see S.I. and dashed lines in Fig.~\ref{Figure2}(c)). 
To avoid a reduction of $I_{\mathrm{c,1}}$ by electron heating due to the dc-power, the bath temperature has to be sufficiently large, allowing the electron-phonon coupling in the heat sinks (see inset of Fig.~\ref{Figure1}) to be effective. In our case we evaluate $\Delta T\approx +1.6$~mK at $T=1.6$~K (see S.I.). 
We follow the amplitude of peaks \textbf{A} and \textbf{B} as a function of the microwave power for two frequencies as shown in figures \ref{Figure2}(d) and (e).
As one increases the power, peak \textbf{A}, related to the first harmonic, decreases following roughly a zeroth order Bessel function (see blue lines in figures \ref{Figure2}(d) and (e)). Peak \textbf{B}, that corresponds to the second harmonic, has a more complicated behaviour. It starts from a non-zero value \cite{note3}, vanishes and then displays a second maximum at higher rf powers and high frequencies in a way whose height is not consistent with the adiabatic phase dynamics (compare yellow squares and lines in figures \ref{Figure2}(d) and (e)).

From the power dependence of the harmonics weight of the CPR obtained from peaks \textbf{A} and \textbf{B}, it is possible to reconstruct, up to a scaling factor, a power-dependent critical current that one may compare to the measured one. To do so, we reconstruct a CPR based on the first two measured harmonics and take its maximum value. The result is reported as dashed lines in figure \ref{Figure2}(a) and demonstrates reasonable agreement with the measured $I_{\textrm{c}}$~\cite{note4}. Such a verification indicates that measuring the ac-Josephson effect for small, but finite, dc-voltage is a good probe of the CPR. This justifies the use of the existing theory of diffusive SNS junctions under microwave irradiation at zero dc voltage.

In the following we use the theory developed by Virtanen \textit{et al.} \cite{Virtanen2010} to account for our experimental data. In this theory both the spectral current density and the out-of-equilibrium distribution function can be obtained by solving the Usadel equation in the  Keldysh-Nambu representation. When $\Gamma<2E_{\mathrm{g}}/\hbar$, the microwave bias affects the distribution function more efficiently than the spectral current density which acquires, however, a component at the frequency of the drive. The dynamics of the current couples back to the distribution function which strongly modifies the CPR. To understand qualitatively the back-coupling of the ac-current to the distribution function we can analytically write the modifications of the distribution function $\delta f=f-f_0$ in the linear response limit. It reads: 
\begin{multline}
\Gamma \left<\rho\right>\delta f=\eta_-(E+\hbar \omega_{\mathrm{rf}}) f_0(E+\hbar \omega_{\mathrm{rf}})[1-f_0(E)]\\
-\eta_+(E) f_0(E)[1-f_0(E+\hbar \omega_{\mathrm{rf}})]\\
+\eta_+(E-\hbar \omega_{\mathrm{rf}}) f_0(E-\hbar \omega_{\mathrm{rf}})[1-f_0(E)]\\
-\eta_-(E) f_0(E)[1-f_0(E-\hbar \omega_{\mathrm{rf}})].
\end{multline}
Here, $\left<\rho\right>$ is the spatially averaged density of states inside the junction. $f_0(E)$ is the equilibrium Fermi-Dirac distribution function and $\eta_+(E)$ and $\eta_-(E)$ are the energy-dependent photon absorption and emission rates, respectively.
At low frequencies $\omega_{\mathrm{rf}}<2E_{\mathrm{g}}/\hbar$, the transition rates are given to a good accuracy by unperturbed spectral functions, similarly as in the Eliashberg \cite{Eliashberg1986} and Mattis-Bardeen \cite{Mattis1958} theories of photoabsorption. At $\omega_{\mathrm{rf}}> 2E_{\mathrm{g}}/\hbar$, however, the ac current flowing in the weak link starts to break Cooper pairs (\textit{i.e.} promote quasiparticles across the gap). An accurate description of the energy dependence of this process requires a more complete consideration of the dynamics of the spectral quantities.

\begin{figure}[htbp]
	\begin{center}
		\includegraphics[width=\linewidth]{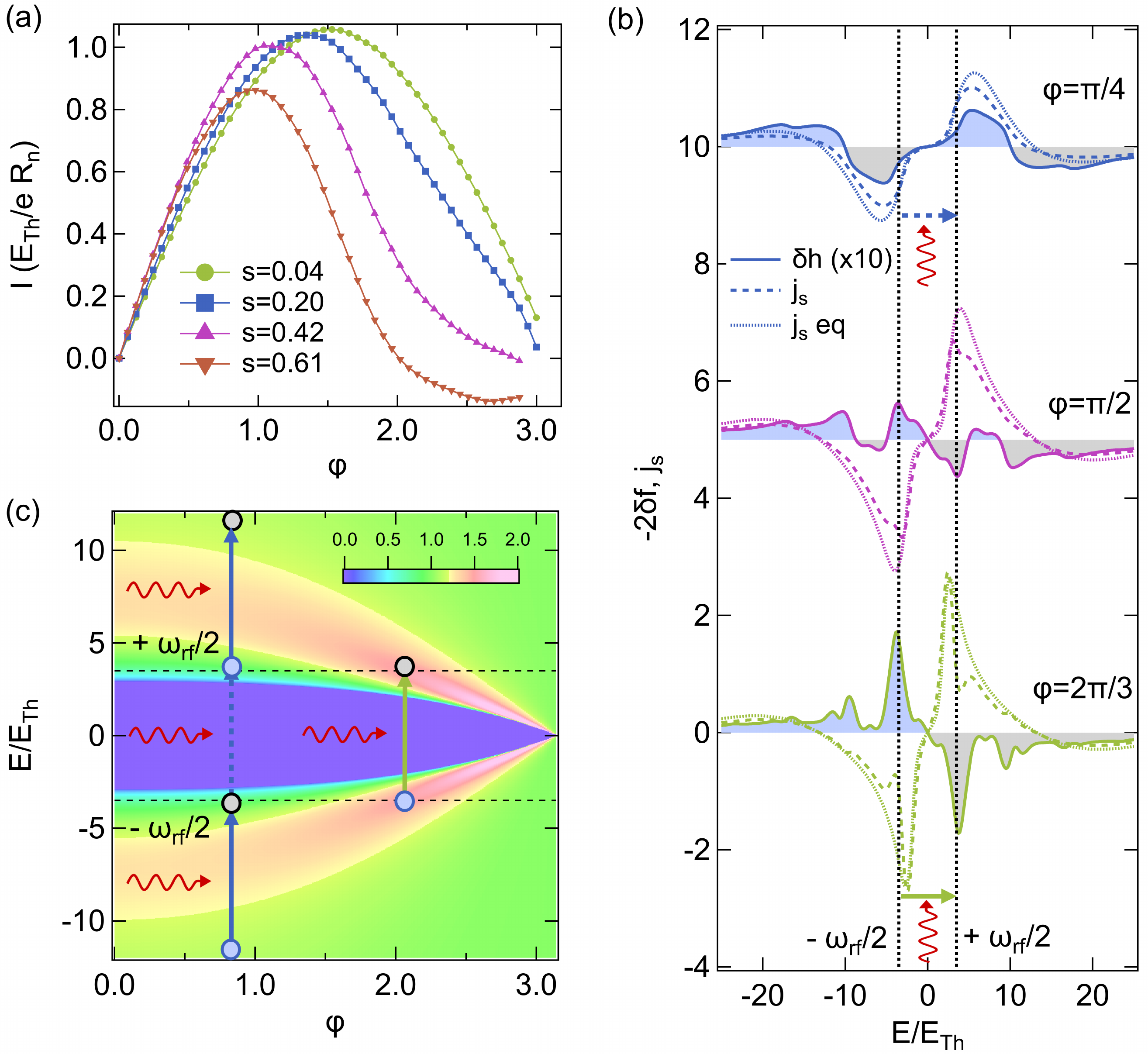}
	\end{center}
	\vspace{-15pt}
	\caption{(a) Calculated current-phase relation for different reduced power $s$. Calculation parameters are $\hbar\omega_{\textrm{rf}}/E_{\mathrm{Th}}=7$, $\Gamma/E_{\mathrm{Th}}=0.4$, $k_{\mathrm{B}} T/E_{\mathrm{Th}}=7$ and $\Delta/E_{\mathrm{Th}}=55$. (b) Calculated equilibrium  $j_{\mathrm{s,eq}}$ and non-equilibrium $j_{\mathrm{s}}$ spectral currents and modifications of the distribution function $-2\delta f$ for different phases. Calculation parameters are as in (a). Full (dashed) horizontal arrows represent the high (low) probability interband transitions. (c) Color-coded sketch of the normalized energy-phase dependent density of states of a long diffusive SNS junction. Full (dashed) vertical arrows represent the high (low) probability inelastic transitions. Grey (blue) circles represent electron-like (hole-like) quasiparticles.}
	\label{Figure3}
\end{figure}

We solve the Usadel equations numerically using the experimental parameters $E_{\mathrm{Th}}$, $\omega_{\mathrm{rf}}$ and the quasiparticle relaxation rate $\Gamma$ close to the above inferred value. We compute the time-average spectral current under the high-frequency drive $\omega_{\mathrm{rf}}$, which yields the effective current-phase relation $I(\varphi,\textrm{s})$ relevant for the lower-frequency phase dynamics (see S.I. and \cite{Virtanen2010}). The result is shown in figure \ref{Figure3}(a)  for the irradiation frequency $\omega_{\mathrm{rf}}/2\pi=35.18$~GHz. As the power is increased, the current-phase relation is distorted and shows a maximum shifted towards smaller phase values. This negative shift demonstrates that the second harmonic value is positive under illumination and not negative as expected from the equilibrium CPR at low temperatures \cite{Heikkila2002}. We quantitatively extract the weights of the different harmonics by fitting the calculated CPR with the formula $I=\sum_{k=0}^{9} I_{\mathrm{c,k}} \textrm{sin} (k \varphi)$ where $I_{\mathrm{c,k}}$ are the fitting parameters. We show in figure \ref{Figure2}(f) and (g) the power dependence of the first two harmonics squared, $I_{\mathrm{c,1}}^2$ and $I_{\mathrm{c,2}}^2$ (Eq.~\ref{CPReq2}), which should be proportional to the experimental spectral density $N_{\textrm{J}}$.

In order to obtain a comparison between the theory and the experiment, at low power, we have to include a negative phenomenological contribution $I_{\rm{c,2~pheno}}$ to match the measured second harmonic at $s=0$. Its precise origin remains to be determined \cite{note3}. In this way, the experimental data shown in figure \ref{Figure2}(d) and (e) coincides with a corrected version of the calculations $I_{\mathrm{c,2}}^2~ \textrm{corr.}=(I_{\mathrm{c,2}}-|I_{\mathrm{c,2~pheno}}| J_{0} (4s))^2$ (see S.I. for details). This correction provides a good agreement between the theory and the experimental data in the full power range with little effect at high power where the strongly non-adiabatic regime appears (see dashed and solid yellow lines around $s\approx0.7$ in Fig.~\ref{Figure2}(f) and (g)). As demonstrated by the purple dashed lines in figure \ref{Figure2}(f) and (g), the Eliashberg theory \cite{Eliashberg1986} fails to explain our experimental data because it neglects the coupling between the phase dynamics and the distribution function.     

The distortion of the CPR can be understood by inspection of the microwave-induced changes of the spectral supercurrent $j_{\mathrm{s}}(E,\varphi)$ and distribution functions $-2\delta f(E,\varphi)=-2[f(E,\varphi)-f_{0}(E)]$ shown in figure \ref{Figure3}(b). For small values of the phase $\varphi$ (see top curves in figure \ref{Figure3}(b)), the changes in the distribution function are dominated by intraband transitions leading to the function $-2\delta f$ and $j_{\mathrm{s}}$ having the same sign and shape. For larger phase values instead, transitions across the gap are favored and visible as peaks in the distribution function (see central and lower curves in figure \ref{Figure3}(b)).
These peaks are located at energies $E=\pm \hbar \omega_{\mathrm{rf}}/2e$, \textit{i.e.}, at the middle of the energy ranges $|E|\in[E_{\textrm{g}}, \hbar\omega_{\textrm{rf}} - E_{\textrm{g}}]$ participating in across-the-gap transitions. Note that the peak positions (vertical lines in Fig.~\ref{Figure3}(c)) are independent of $E_{\textrm{g}}$. The peaks originate from the transition probability that is influenced by the ac response of the spectral supercurrent, which deviates from the equilibrium one as shown in figure \ref{Figure3}(c). Importantly, these peaks have a sign that is opposite to the spectral current implying that the Cooper pair breaking results in a reduction of the total supercurrent.

In conclusion, we performed a microwave spectroscopy of the ac-Josephson effect in a diffusive weak link in the \textit{strongly non-adiabatic} regime for which inelastic transitions accross the minigap are possible. The microwaves are found to drastically enhance the second harmonic of the CPR as a result of the back-coupling of the ac-spectral supercurrent to the distribution function.
Future experiments shall investigate the Josephson emission at high frequency in limits where the frequency of the emitted photons is comparable to the minigap in the normal wire \cite{Riedel1964,Hamilton1971}. Besides diffusive-metal SNS junctions, the spectroscopic approach could be used for several other types of weak links. In particular, microwaves also modify the CPR in atomic contacts \cite{Bergeret2010,Bretheau2013}. In nanowire junctions with Majorana bound states, the microwave affected CPR might reveal signatures about the topologically forbidden transitions \cite{Deacon2017,Ren2019,Fornieri2019}.
	
We acknowledge valuable discussions with B. Reulet, F. Massee, H. Bouchiat, M. Ferrier, R. Deblock and S. Gu\'eron. This work has partially been funded by the European Union's Horizon 2020 research and innovation programme under grant agreement No 800923, and the Academy of Finland grant number 317118. C. Strunk thanks the CNRS and the Universit\'e Paris-Sud for funding his stay at the Laboratoire de Physique des Solides.


\begin{thebibliography}{99}
\bibitem{note1} If the energy relaxation rate is the larger energy scale, the junction displays microwave-induced phase dynamics only, as in tunnel junctions.
\bibitem{note0} The measured critical current is the current at which the differential resistance reaches the trigger value dV/dI$=0.04$~$\Omega$. This current might differ from the intrinsic critical current by a multiplication factor (see S.I.) due to current fluctuations. The microwave power dependence which is of interest here is however qualitatively unaffected.
\bibitem{note2} The values of $N_{\textrm{J}}$ take into account the whole microwave setup calibration. A systematic scaling uncertainty is then expected in the extracted values. See S.I. for the noise spectral density calibration.
\end{thebibliography}
\end{document}